\documentclass[11pt,twocolumn,a4paper]{article}
\usepackage[american]{babel}
\usepackage[utf8]{inputenc}
\usepackage{subcaption}
\usepackage{graphicx}
\usepackage[export]{adjustbox}
\usepackage{dsfont}
\usepackage{amsmath}
\usepackage{amssymb}
\usepackage{tabularx}
\usepackage{multirow}
\usepackage[table]{xcolor}
\usepackage{float}% If comment this, figure moves to Page 2
\usepackage{flushend}
\usepackage{appendix}

\usepackage{multicol}
\usepackage{titlesec}
\titleformat{\section}
  {\normalfont\fontsize{12}{15}\bfseries}{\thesection}{1em}{}
\usepackage{titlesec}
\usepackage{authblk}
\usepackage{lipsum}
\usepackage{cite}
\usepackage{cleveref}
\usepackage{layout}
\usepackage{ulem}
\begin{document}
\title{Sampling $U(1)$ Gauge Theory using a re-trainable Conditional Flow-based model}

\author[1]{Ankur Singha}
\author[1]{Dipankar Chakrabarti}
\author[2]{Vipul Arora}
\affil[1]{Department of Physics, Indian Institute of Technology Kanpur, Kanpur-208016, India}
\affil[2]{Department of Electrical Engineering, Indian Institute of Technology Kanpur, Kanpur-208016, India}
\date{}
\twocolumn[
  \begin{@twocolumnfalse}
    \maketitle
\begin{abstract}
Sampling topological quantities in the Monte Carlo simulation of Lattice Gauge Theory becomes challenging as we approach the continuum limit ($a \rightarrow0$) of the theory. In this work, we introduce a Conditional Normalizing Flow (C-NF) model to sample $U(1)$ gauge theory in two dimensions, aiming to mitigate the impact of topological freezing when dealing with smaller values of the $U(1)$ bare coupling. To train the conditional flow model, we utilize samples generated by Hybrid Monte Carlo (HMC) method, ensuring that the autocorrelation in topological quantities remains low. Subsequently, we employ the trained model to extrapolate the coupling parameter to values where training was not performed. We thoroughly examine the quality of the model in this region and generate uncorrelated samples, significantly reducing the occurrence of topological freezing. Furthermore, we propose a re-trainable approach that utilizes the model's own samples to enhance the generalization capability of the conditional model. This method enables sampling for coupling values far beyond the initial training region, expanding the model's applicability.
\end{abstract}
  \end{@twocolumnfalse}
  ]

\section{Introduction}
In the field of Lattice theory, Monte Carlo simulation methods are utilized to sample configurations of the lattice. These samples are generated based on a distribution determined by the action of the lattice theory. The choice of the action parameter value used in generating the lattice samples determines the computational cost of the simulation. As we approach the critical region of a lattice system or move towards the continuum limit of a lattice field theory, the samples become highly correlated. Within the critical region, the integrated autocorrelation time, which measures the level of correlation, increases rapidly and diverges at the critical point. In a finite-size lattice, the critical point corresponds to the peak of the autocorrelation curve, leading to a phenomenon known as critical slowing down \cite{Wolff:1989wq,Schaefer:2010hu}. This has been a major problem in the simulation of lattice systems near the critical region. If proper care is not taken this may lead to biased estimation of observable.

Monte Carlo simulation encounters similar difficulties when exploring the topological sector in lattice Gauge theory, such as $ U(1)$ gauge theory in 2D. When we move towards finer lattices, the topological charge freezes, i.e., the HMC sampling is restricted only to one or two topological sectors for a longer simulation run. This implies a larger integrated autocorrelation time for topological charge which is well known as topological freezing. Numerous efforts have been made to mitigate the impact of critical slowing down in statistical systems and lattice QFT \cite{Ramos:2012bb, gambhir2015improved,Endres_2015}. However,  in lattice gauge theory it still remains a challenging task.

These days, ML-based solutions to the critical slowing down problem has become popular. Various ML algorithms have been applied for statistical physics ,condensed matter problems and Lattice Field Theory\cite{Albandea:2021lvl,Zhou:2023pti,Lehner:2023prf,Aronsson:2023rli,Nicoli:2023qsl,Bacchio:2022vje,PhysRevD.100.011501, PhysRevLett.122.080602,Pawlowski_2020,Nicoli_2020,PhysRevB.94.195105,Singh_2021,Carrasquilla_2020, Liu_2017, Vielhaben_2021, Chen_2018,PhysRevB.94.165134,Science.355.602,PhysRevLett.120.066401,Abbott:2023thq,Coccaro:2023vtb,Abbott:2022hkm,Abbott:2022zhs,Favoni:2022mcg}. 
 Several generative learning algorithms have recently been developed to avoid the difficulty in lattice field theory\cite{PhysRevD.100.034515,albergoflowbased,shanahan2018machine,Kanwar_2020,albergo2021introduction,Albergo:2022qfi,Hackett:2021idh,deHaan:2021erb,Caselle:2022esc,Caselle:2022acb,Caselle:2023mvh,R:2023dcr}. In the flow-based approach \cite{PhysRevD.100.034515,Kanwar_2020}  NF models are trained for a action parameter with reverse KL divergence. 
But this kind of self-learning method has a major issue of mode collapse which may lead to inefficient modeling of complicated multi-modal distributions\cite{Hackett:2021idh,Nicoli:2023qsl}. In contrast a generative model trained with forward KL are mode covering and applying MH can produce the correct statistics. In the U(1) gauge theory, this issue of mode collapse is discussed in \Cref{mode_colapse}. In our earlier works \cite{Singha:2023cql,Singha:2021nht} we  presented a method for sampling lattice configurations near the critical regions using Conditional Normalizing Flow (C-NF) (conditional normalizing flow has also recently been used in\cite{pawlowski2022flowbased})and Conditional GAN to reduce the problem of critical slowing down for Scalar field theory and lattice Gross-Neveu model in 2 dimensions. We have shown that the C-NF model \cite{Singha:2023cql} trained in the non-critical region can produce samples for parameter values in the critical region. 
In this work, we propose an application of equivariant flow to construct a conditional flow model for sampling $U(1)$ gauge theory in 2 dimensions. To the best of our knowledge the conditional flow model, based on the action parameter, has not been employed in the sampling of lattice gauge theory. 

\begin{figure}[ht!]
    \begin{subfigure}{.5\textwidth}
      \includegraphics[width=\textwidth]{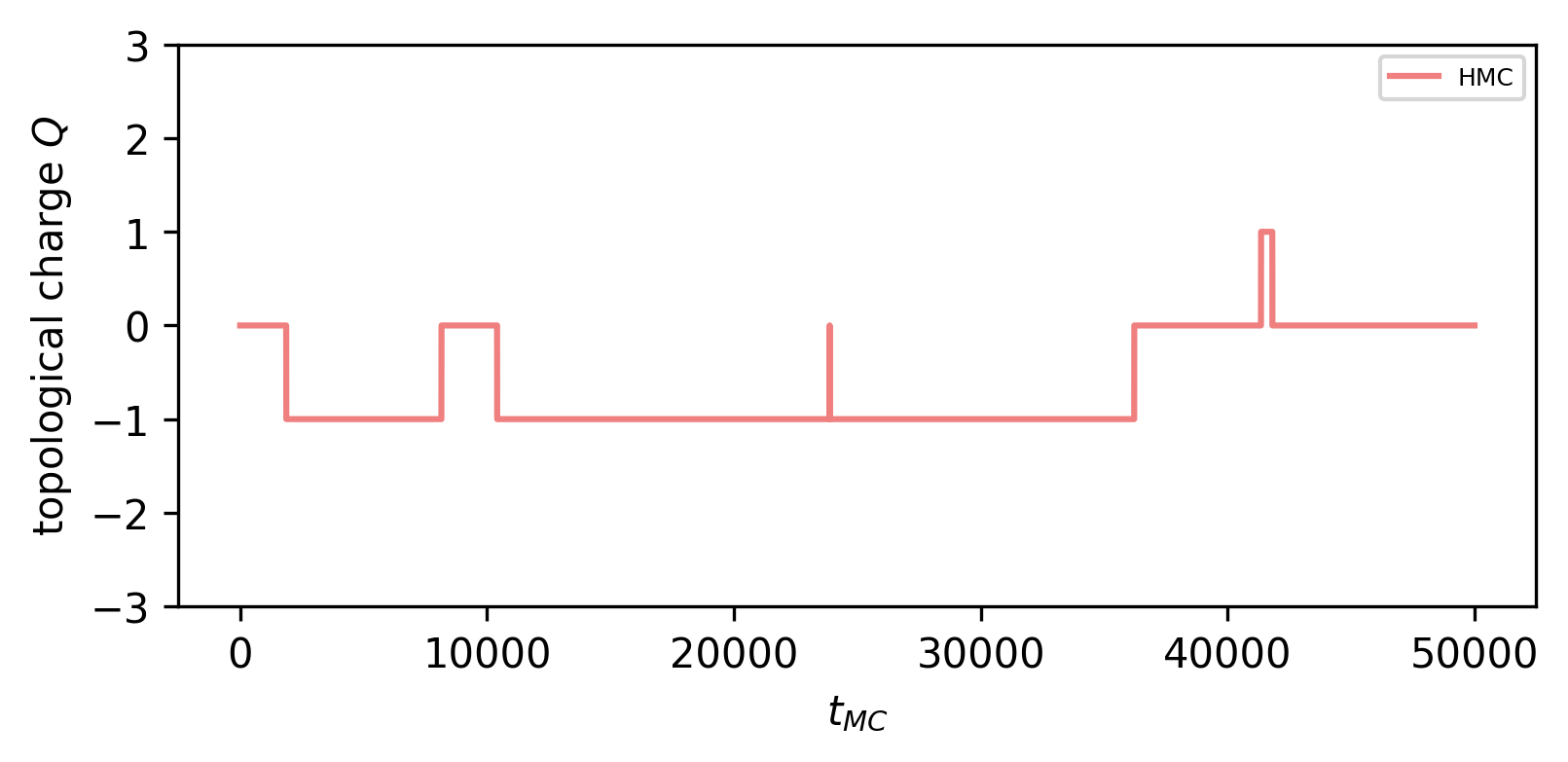}
      \caption{$\beta=7$}
      \label{tp_intro}
    \end{subfigure}
    \begin{subfigure}{.5\textwidth}
        \includegraphics[width=\textwidth]{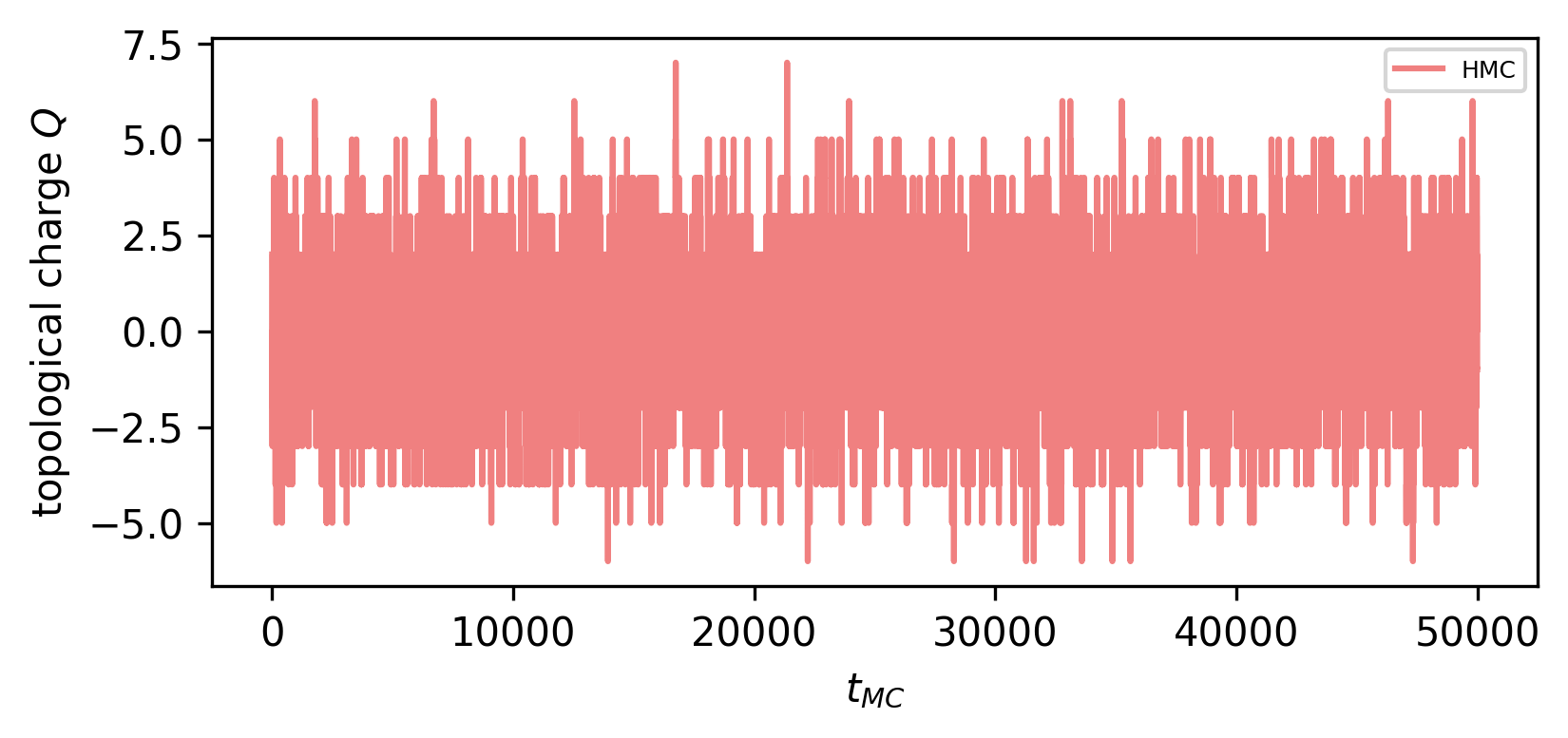}
        \caption{$\beta=3.5$}
        \label{tp_datset}
    \end{subfigure}
\caption{The fluctuation of topological charge for $50k$ Markov steps simulated by HMC at (a) $\beta=7$ and (b) $\beta=3.5$.}
\label{corr_ext}
\end{figure}

In U(1) gauge theory, the target distribution for a given action parameter $\beta$ can be defined as,
\begin{align}
p(U_\mu(n)|\beta) =\frac{1}{Z} e^{-S(U,\beta)},
\label{distr} 
\end{align}
where $U_\mu(n)$ denotes the $U(1)$ Link variable field, $\beta$ denotes the lattice action parameter and Z is the partition function defined as $ Z=\sum\limits_{U}e^{-S(U,\beta)}$.
The action parameter $\beta$ is the conditional parameter, and we train a conditional flow model on lattice ensembles corresponding to different $\beta$ values. 
We partition the action parameter into two sets
$\beta_L$ and $\beta_S$ based on the integrated autocorrelation time ($\tau_{int}$) of topological charge. $\beta_L$ corresponds to large $\beta$ values where topological freezing is dominant in HMC simulations as shown in \Cref{tp_intro} and $\beta_S$ corresponds to smaller values of $\beta$ where the $\tau_{int}$ is small, and hence the samples fluctuate among different topological sectors as shown in \Cref{tp_datset}. 
Due of the lower autocorrelation time, the HMC simulation cost is minimal in the $\beta_S$ set. Hence we train a Conditional Flow Model (C-NF) $\tilde{p}({U_\beta})$ with HMC samples from $p(U,\beta_{S})$.

 We train the C-NF model to be a generalised model over $\beta$ parameters. The model is then extrapolated to larger values of $\beta $, i.e in the $\beta_L$ set to generate different ensembles of lattice configurations. However, the extrapolated model may not provide samples from the true target distribution. But the exactness can be guaranteed by using the MH algorithm. Therefore, we use the extrapolated model $\tilde{p}(U{(\beta_{L})})$ at large $\beta$ values as proposal for constructing a Markov Chain via an independent MH algorithm\cite{PhysRevD.100.034515}. The quality of the extrapolated model may deteriorate as we move further away from the training region, particularly for distant $\beta$ values. In such cases, we can employ a re-trainable method to sample at those $\beta$ values. This re-trainable method utilizes the model's own samples at intermediate $\beta$ value to enhance the generalization of the C-NF model, enabling better sampling capabilities at points that are far away from the training set. 
 \section{$U(1)$ Gauge Theory}
 The lattice action for $U(1)$ gauge theory in $1+1$ dimensions can be written as
 \begin{align}
        S(U)=-\beta \sum\limits_{n,\mu<\nu}Re[U_{\mu\nu}(n)],
 \end{align}
where, the plaquette $U_{\mu\nu}$ is defined as
 \begin{align}
     U_{\mu\nu}= U_\mu(n)U_\nu(n+\hat\mu)U_\mu(n+\hat\nu)U_\nu(n).
 \end{align}
$U_\mu(n)$ can be written in terms of angular variable as
$U_\mu(n)=e^{i\theta_\mu(n)}$.
The action is symmetric under the transformation
\begin{align}
U_\mu(n) \longrightarrow e^{i\alpha(n)}U_\mu(n)e^{-i\alpha(n+\mu)}.
\end{align}
In terms of angular variable the plaquette becomes
\begin{align}
  U_{\mu\nu}(n)=&e^{i[\theta_\mu(n)+\theta_\nu(n+\hat\mu)-\theta_\nu(n+\hat\nu)-\theta_\nu(n)]}, \\
  =& e^{i\theta_{\mu\nu}}(n).
\end{align}

So, the action can be written as
\begin{align}
    S(U)=-\beta\sum\limits_{n} \cos{\theta_{\mu\nu}(n)},
    \label{act}
\end{align}

which is symmetric under
\begin{align}
    \theta_\mu(n)\longrightarrow\alpha(n) +\theta_\mu(n)-\alpha({n+\hat\mu}),\\~~ \text{where}~~ \alpha({n})\in R \notag
\end{align}
The action in \Cref{act} is used in \Cref{distr} for sampling in both HMC and C-NF model.
The observable we calculate here is the topological charge defined as
\begin{align}
    Q=\frac{1}{2\pi}\sum\limits_{n}arg [U_{\mu\nu}(n)], \\
    \text{where},~~ arg[(U_{\mu\nu}]\in [-\pi,\pi]).
\end{align}

\section{C-NF Model employing Equivariant Flow} In a flow-based model, starting with a gauge symmetric prior distribution does not guarantee that the resulting output distribution will also be gauge symmetric. For the output distribution to maintain gauge symmetry, the flow transformation must be an equivariant transformation, meaning it should commute with the gauge transformation. Implementing such equivariance in a flow-based model is non-trivial. However, if the flow transformation only affects the gauge invariant quantity, then the condition mentioned above is easily satisfied. In such cases, the flow-based model can maintain gauge symmetry by exclusively transforming the gauge invariant component. In $U(1)$ gauge theory, the link variable $U_{\mu}(n)$ are not itself gauge invariant. For such construction one has to find one to one correspondence between the gauge invariant quantity and the link variable. For example, plaquette variable $P_{\mu\nu}(n)$ is one such gauge invariant quantity which can be used for construction of equivariant flow. We use the equivariant flow construction following Ref.\cite{Kanwar_2020}. In 2D, the link variable are tensor of shape $(2,L,L)$ and  the plaquette variable are tensor of shape $(L,L)$. The update in $P_{\mu\nu}$ can be translated into update of a single $U_\mu$ using suitable mask pattern. Note that a single update of a link variable in 2D will update two adjacent plaquttes. Therefore, in a coupling layer we need three subclass of input plaquettes as described below, obtained by suitable mask namely active, passive and frozen \cite{Kanwar_2020}. The active variables $(P^a_{\mu\nu})$ are transformed by the coupling layer, the frozen variable $(P^f_{\mu\nu})$ are not transformed by this coupling layer and the passive variables $(P^{ps}_{\mu\nu})$ are those which gets transformed indirectly due to a link update. We choose the mask such that the update in the link variable $U_{\mu\nu}$ is sparse and hence scalable to larger lattice sizes. We use a total $32$ coupling layers in the C-NF model.

 The link update in a coupling layer happens in two steps, in the first step, we take care of translating $P_{\mu\nu}$ to $U_\mu$ and the second step corresponds to the actual flow in the $P_{\mu\nu}(n)$ variable. For the second step, we use the Non-Compact Projection (NCP) flow \cite{Rezende:2020hrd}. In NCP flow, we use fully convolutional layers to transform the frozen variables of the coupling layers. We have used $48$ such coupling layers, and each coupling layer has three hidden CNNs with a number of filters $16,32,16$. We use LeakyReLU activation in the hidden layers. Since we use coupling layers the conditional parameters can be concatenated to frozen variables and passed through the neural network, which is evaluated in the forward direction only.

\section{Numerical Experiments and Results} In this section, we present the details of our numerical experiment and the resulting outcomes. We outline the training and sampling processes and the dataset preparation used to train the C-NF model. 
%%%%%%%%%%%%%%%%%%%%%%%%%%%%%%

\subsection{Mode Collapse in U(1) gauge theory}\label{mode_colapse}
%%%%%%%%%%%%%%%%%%%%%%%%%%%%%%%%%%%%%%%%%%%%%%%%%%%%%%%%%%%%%%%%%%%%
In the literature, there are several flow-based works which try to accelerate the sampling  of lattice field theory using Reverse KL (RKL) divergence. However, one major disadvantage  RKL  comes with is the mode collapsing behaviour. On the other hand, flow-based models trained using Forward KL (FKL) are mode covering. In this section, we  investigate the problem of mode collapse in a flow-based model for U(1) gauge theory. We train two flow-based models $q_R(U,\Theta)$ and $q_F(U,\Psi)$ using RKL and FKL divergence, respectively. After training both models to  $50\%$ ESS, we estimate the  Negative Log-Likelihood (NLL). Two kinds of NLL for the flow models are used for the investigation, namely $\mathcal{L}_1$ and $\mathcal{L}_2$:
\begin{align}
    \mathcal{L}_1=E_{U\sim p(U|\beta)} [\log q(U,\Theta )],
\end{align}
where $q(U,\Theta)$ is either trained with the FKL or RKL and the expectation $E$ is taken over samples generated from the true distribution (e.g. HMC simulation). This type of NLL is best suited for detecting mode collapse in a generative model. 
\begin{align}
   \mathcal{L}_2=E_{U\sim q_U,\Theta) } [\log p(U|\beta)] 
\end{align}
In this case, the expectation is taken over the samples generated from the model itself and the expectation of $p(U)$ is estimated. This metric alone is not sufficient  to investigate the mode collapse as explained in the appendix. However, by the combination of $\mathcal{L}_1 $ and $\mathcal{L}_2$, one can detect the mode collapse.

For HMC $\mathcal{L}_1=\mathcal{L}_2$  and is given by
\begin{align}
    \mathcal{L}_{hmc}=E_{x\sim p(U|\beta) }[\log p(U\beta)]
\end{align}

\begin{table}
\begin{tabular}{ |p{1cm}||p{2cm}|p{2cm}| }

\hline
\multicolumn{3}{|c|}{NLL comparison} \\
\hline
Model & $\mathcal{L}_1$ & $\mathcal{L}_2$  \\
\hline
RKL\cite{albergo2021introduction} & 270.611 & 196.882   \\
\hline
FKL & 199.715 & 196.982   \\
\hline
HMC & \multicolumn{2}{c|}{196.666} \\
\hline
\end{tabular}
\caption{ Comparision of the different NLL types. For HMC $\mathcal{L}_1$ and $\mathcal{L}_2$ are equal. }
\label{mode_tab}
\end{table}
We have estimated all three NLL types, $\mathcal{L}_1,~\mathcal{L}_2 $ and $\mathcal{L}_{hmc}$ shown in the \Cref{mode_tab}. We see that the $\mathcal{L}_1$ is quite high for the RKL  and highly deviates from the $\mathcal{L}_{hmc}$. For RKL, $\mathcal{L}_1$ is 270.61, which is quite higher than the $\mathcal{L}_{2}$ (196.88). This indicates that the RKL model has not explored the distribution's parameter space as HMC does. Hence, RKL generates samples in a specific region of the distribution, leading to the disagreement in $\mathcal{L}_{2}$ and $\mathcal{L}_{1}$. On the other hand, for the FKL, $\mathcal{L}_1$ is low and very close to the $\mathcal{L}_{hmc}$. Moreover, $\mathcal{L}_{1}$ and $\mathcal{L}_{2}$ are also quite close (196.98 and 199.71) for FKL. More discussions on this issue are given in the Appendix.

Note that an RKL model alone is sufficient to observe the mode collapse, and one can use the method \cite{albergo2021introduction} for $U(1)$ gauge theory to check the large disagreement of $\mathcal{L}_{1}$ and $\mathcal{L}_{2}$. The observed difference remains constant while training, even if we increase the ESS. This is what one expects in an online training process.
%More discussion on  $\mathcal{L}_{1}$ and $\mathcal{L}_{2}$ is given in  \Cref{App_a}.
 
\subsection{Training dataset}
To train the C-NF model, we generate $10$ different ensembles of lattice configurations, each corresponding to a different value of $\beta=\beta_S:\{1.0,1.5,1.8,2,2.2,2.5,2.8,3,3.2,3.5\}$. We choose the training dataset such that there is no topological freezing. The largest value of $\beta_S$ is $3.5$, where integrated autocorrelation time is $\approx 23.45$. The behaviour of topological charge is shown in the \Cref{tp_datset} for $\beta_S=3.5$. For larger $\beta$ the topological freezing sets in as shown in \Cref{tp_intro}. To prepare training ensembles, we employ HMC simulation and adjust the HMC parameter for each $\beta_S$ value to achieve an acceptance rate of approximately $ \approx 85\%$. We have used a non-uniform ensemble size for each $\beta_S$. For the largest value of $\beta_S$, we generate $15000$ samples and as we move to the next lower $\beta_S$ we reduce ensemble size by $500$. We perform all the numerical experiments on a $16\times16$ lattice.

%%%%%%%%%%%%%%%%%%%%%%%%%%%%%%%%%
\subsection{Training and sampling}
In the C-NF model, the input lattice configurations are tensors with a shape of $(2, 16, 16)$. These tensors are concatenated with their corresponding ensemble labels or conditional parameters, denoted as $ \mathds{1}\beta$. The tensor $\mathds{1}$ has the same shape as the lattice configurations, $(2, 16, 16)$, with all elements set to identity. It implies that all masking patterns will be applied to the condition tensors along with the lattice configurations.

During training, we randomly select a batch from any value of $\beta_S$. We use a batch size of $512$ to calculate the gradient at each iteration, and the model weights are updated after every $10$ iteration. The C-NF model is trained using forward KL divergence and an Adam optimizer with an initial learning rate of $0.001$.
Increasing the Effective Sample Size (ESS) of the C-NF model can be challenging and may reach a plateau during the training. We incorporate a decay of learning rate of $0.5$ at intervals of $25000$ training iterations. We have found this approach effective in increasing the ESS for the C-NF model.
 It's important to note that the C-NF model may exhibit overfitting and perform well only for the training $\beta$ values. To overcome this, we condition only every fourth coupling layer allowing for better generalization and performance on unseen $\beta$ values.
  We stop the training when the ESS reach above $30\%$ and the increment is less than $3\%$ for next $5,000$ consecutive iterations. Note that, with further training or a more optimized architecture, one can potentially achieve a higher ESS. However, our objective is to assess the sample quality as we extrapolate for large $\beta$ values. While training, we also monitor the acceptance rate periodically after every $10,000$ iterations in the MH algorithm. At the end of training, we achieve an acceptance rate of approximately $65\%$ in the training region.

After training, we extrapolate the model for large $\beta$ values, $\beta_L=\{5.5,6,6.5,7,7.5\}$. Using the extrapolated model, we generate proposals for the MH step in order to construct a Markov chain. For each value of $\beta$, we generate an ensemble consisting of $10^6$ lattice configurations. Once training is over, the flow-based model enables us to efficiently generate such large ensembles without any significant challenges. 

%%%%%%%%%%%%%%%%%%%%%%%%%%%%
\subsection{Results}
\begin{figure}[ht!]
\centering
  \includegraphics[width=.49\textwidth]{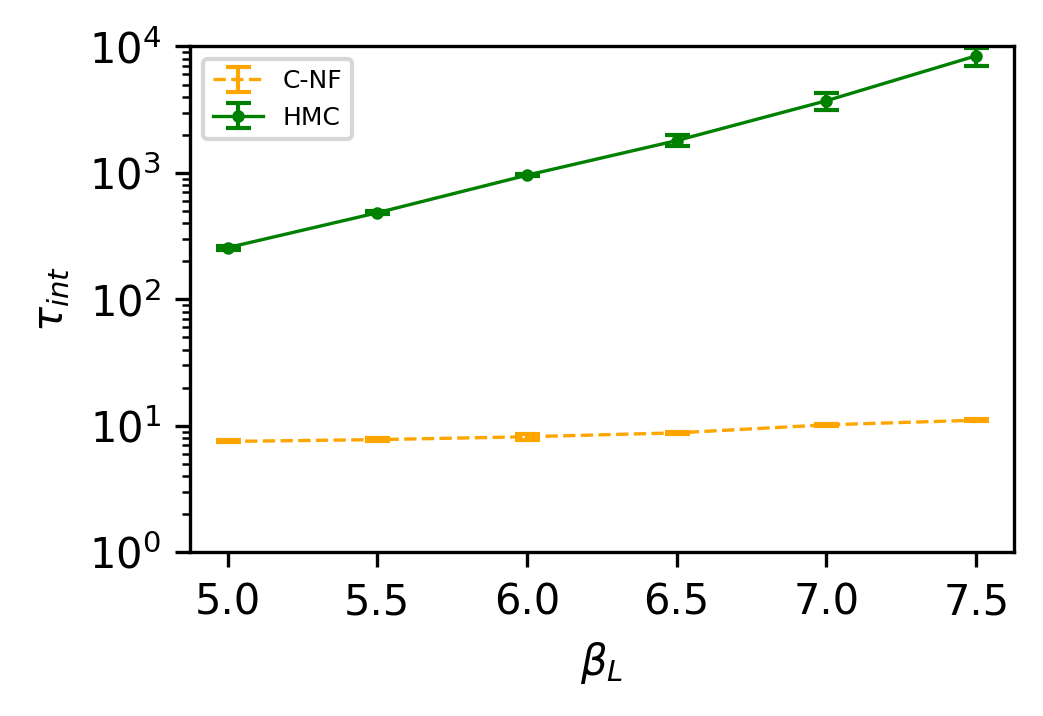}
      \caption{Integrated autocorrelation time calculated form the C-NF model and HMC simulation. The solid (green) line represents $\tau_{int}$ in HMC simulation and  dashed (orange) line represents $\tau_{int}$ in C-NF.}
      \label{compare_int}
\end{figure}

We calculate different observables on the ensemble generated from the C-NF model. In \Cref{compare_int} we have shown the integrated autocorrelation time for HMC and C-NF model calculated for a topological charge. During the estimation of observables on the ensemble, we consider every $20th$ configuration to reduce any effect of autocorrelation. We see that for HMC simulation, the effect of topological freezing increases rapidly as we move towards larger $\beta$. Since the generation in the C-NF model is inherently uncorrelated, the autocorrelation depends on the acceptance rate in MH step. There is a massive gain as we move towards larger $\beta$ where we have not trained the model.  
\begin{figure}[ht!]
\centering
  \includegraphics[width=.49\textwidth]{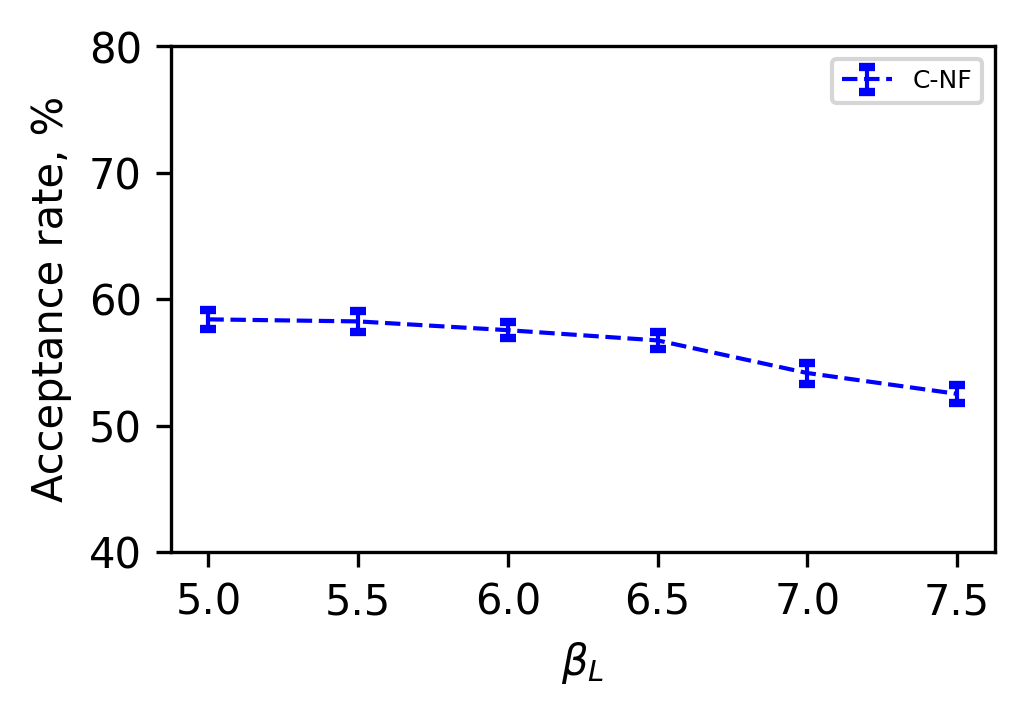}
      \caption{Acceptance rate calculated from the extrapolated C-NF model for different $\beta$ values. The acceptance rate shows a very small decline over a wide range of non-training values of $\beta$.}
      \label{accept}
\end{figure}
In the extrapolated region, the acceptance rate in the MH step is almost constant. This indicates that our model has learned the conditional distribution very well. In \Cref{accept}, we show that the acceptance rate is between $60\%$ to $50\%$ for the extrapolated  parameter values in $\beta_L$. 
%Here, we train a single model for $\beta_S$ and generate samples for different $\beta$ in $\beta_L$.
This allows one to store a single model and generate ensembles at multiple $\beta$ values.

\begin{figure}[ht!]
\centering
  \includegraphics[width=.49\textwidth]{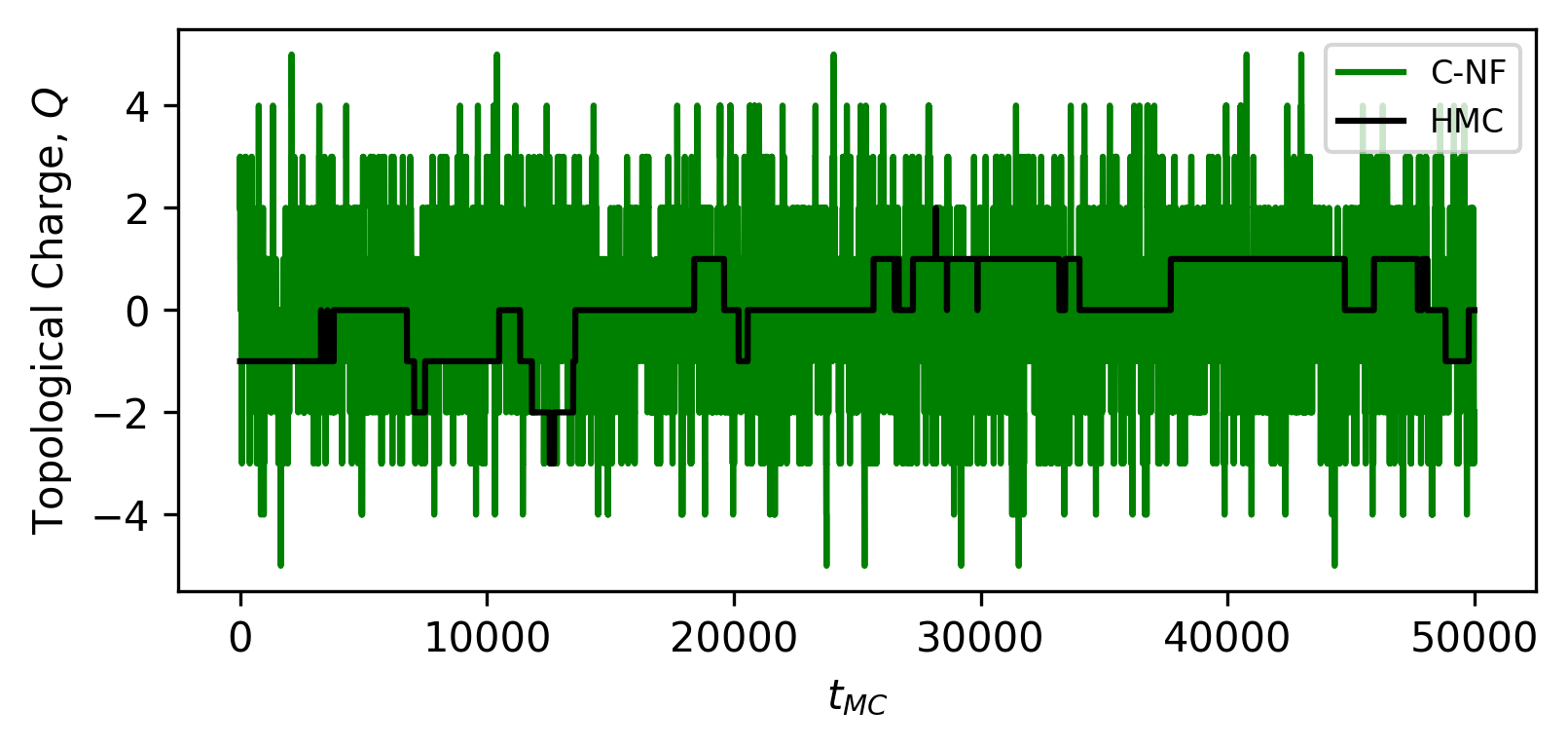}
      \caption{Topological freezing is shown for $50k$ Markov chain in both HMC simulation and C-NF model. A significant gain from the C-NF model in reducing the effect of topological freezing is clearly visible.}
      \label{tp_compare_re}
\end{figure}

\Cref{tp_compare_re} shows the fluctuation of topological charge with the MCMC time. We see clear freezing in the HMC Markov chain. In contrast, the C-NF model has reduced it significantly. This is a significant gain one can obtain from a flow-based model.

\subsection{Re-training method for distant $\beta$ values} If we want to extrapolate the model far away from the training region, the acceptance rate may decrease. This is due to the lack of good generalization for the conditional model over a wide range of $\beta$ values. For example, if we want to generate samples for $\beta=9$, then the MH acceptance rate drops to $\sim 30\%$. We address this issue by utilizing samples from non-training regions of $\beta$ values. For extrapolation at $\beta=9$, first, we generate samples from the C-NF model for an intermediate $\beta$, say $\beta=6$, where the acceptance rate is $\sim 55\%$. We add these newly generated samples and re-train the C-NF model using Forward KL for 10k iterations along with the previously used samples. We reduce $50\%$ samples for each $\beta\in \beta_{S}$. This will improve the generalization of the C-NF model for larger $\beta$ values.

We observed an increase in acceptance rate with the re-training up to $\sim 40\%$ for $\beta=9$.  Note that if $\beta$ is not too far from the training region and  the ESS does not fall off drastically, re-training is not required.

\section{Summary and Conclusion} 
Based on our previous work, it has been observed that the conditional flow model has proven to be successful when applied to Scalar and Gross-Nevue models. Building and Testing the conditional flow-based model on gauge theory is crucial to advance further in the development of lattice QCD.

 In the simulation of $U(1)$ gauge theory using the Monte Carlo Markov Chain (MCMC) method, the correlation between the samples becomes stronger as lattice spacing decrease i.e., $\beta$ increases. As $\beta$ reaches higher values, a phenomenon called topological freezing is observed in the MCMC simulation. To tackle this issue, we have developed a conditional flow-based model (C-NF) specifically designed for sampling the $U(1)$ gauge theory in 2D when $\beta$ is large. This model takes $\beta$ as a conditioning parameter and is trained using ensembles of $\beta$ values where there are low autocorrelations and minimal effects of topological freezing. The training process continues until the Effective Sample Size (ESS) for the training ensembles reaches approximately $30\%$.

Once the training is completed, we employ the model to generate samples for larger $\beta$ values using the Metropolis-Hastings (MH) algorithm. The quality of the extrapolated model depends on how well it has learned a generalized distribution over the $\beta$ parameter. We achieve an acceptance rate of approximately $50-60\%$ in the MH algorithm across a wide range of $\beta$ values that were not included in the training data. For $\beta = 7.5$, the acceptance rate in the MH algorithm was approximately $52\%$. However, as $\beta$ increases further, the acceptance rate gradually decreases. In such cases, if necessary, one can use a re-trainable method for sampling points far beyond the training region. This method utilizes the model's samples at intermediate $\beta$ values to improve the conditional generation of samples.

While the conditional flow model has shown effectiveness in certain cases, it is generally not scalable for sampling high-dimensional distributions. One possible future direction is to tackle the challenging task of constructing a conditional flow model for sampling high-dimensional lattice gauge theory.
\section*{Acknowledgement}
SERB supports this work under grant no. CRG/2021/003466.
\bibliographystyle{unsrt}
\bibliography{name}

\Large Appendix
\section*{Mode Collapse in a Flow model}\label{App_a}

\begin{figure}[ht!]
\centering
  \includegraphics[width=.49\textwidth]{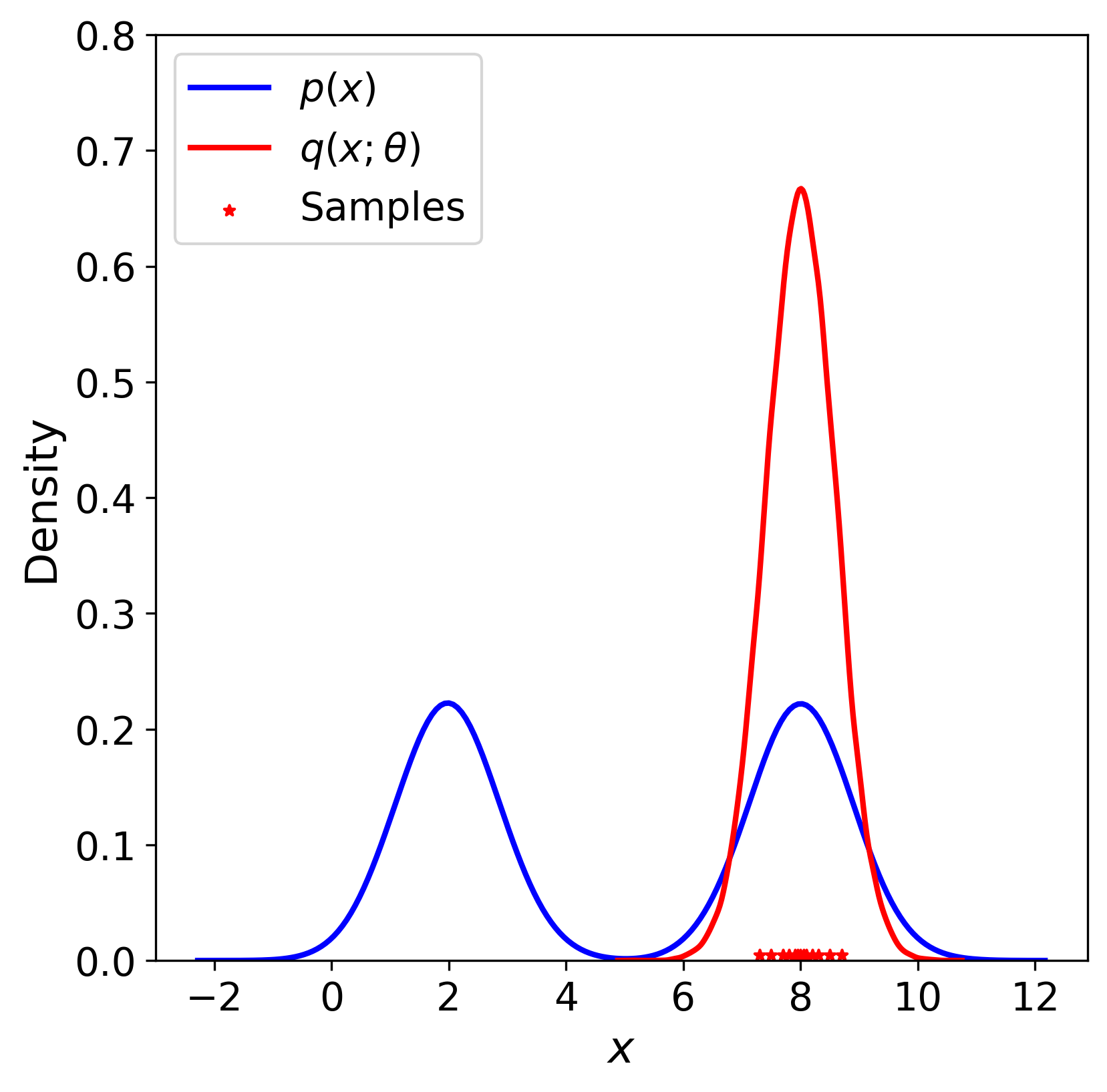}
      \caption{Mode collapse in a Flow model.}
      \label{mode_col}
\end{figure}

Let's examine a multi-modal target distribution, denoted as $p(x)$, which encompasses two modes as illustrated in \Cref{mode_col}. To represent this distribution, we employ an NF model denoted as $q(x,\Theta)$. This NF model can be trained using either forward KL or reverse KL divergence.

We examine two types of NLL metrics to investigate Mode Collapse in the NF model, aiming to determine which one is more effective in identifying a mode collapse. We use HMC as our baseline for studying the mode collapse phenomenon.
 
In $\mathcal{L}2$, samples are drawn from the distribution $p(x)$, where $x\sim p(x)$, and the density is subsequently estimated using the NF model. In other words, we are estimating $E{x\sim p(x)}[\log q(x,\Theta)]$.

When there is no mode collapse, the model generates samples in both modes and as a result, $E_{x\sim p(x)}[\log q(x,\Theta)]$ aligns with $\mathcal{L}_{hmc}$.

Now, consider the scenario in which mode collapse occurs. Since the model learns only a single mode, $q(x,\Theta)$ becomes substantially elevated at that specific mode, as illustrated in \Cref{mode_col}. Consequently, $\mathcal{L}_1$ will be notably higher compared to the situation where all modes are learned. Therefore, this serves as a suitable metric for identifying mode collapse.

In $\mathcal{L}_2 $, samples are generated from the trained NF model and the expectation value of $\log p(x)$ is estimated. 
If we assume there is no mode collapse, then the model will generate samples $x\sim q(x)$ in both modes and $E_{x}[\log p(x)]$ will match with $\mathcal{L}_{hmc}$.
If there is a mode collapse, the model will generate samples only in a single mode. But the estimate $E_{x}[\log p(x)]$ will still match with $\mathcal{L}_{hmc}$ as the estimated quantity is $\log p(x)$. Thus it does not serve as a good metric for mode collapse detection.

\end{document}